\documentclass[11pt]{article}
\usepackage{graphicx}
\usepackage{amssymb}
\usepackage{amsmath}
\usepackage{graphicx}
\usepackage{amstext}
\usepackage{esint}
\usepackage[utf8]{inputenc}
\usepackage{color}
\usepackage{cite}

\renewcommand{\vec}[1]{{\bf #1}}       
\def\beq{\begin{eqnarray}}    
\def\eeq{\end{eqnarray}}      

\newcommand{\rL}{\rho_\Lambda}

\newcommand{\CC}{\Lambda}

\newcommand{\Omo}{\Omega_{m}}

\newcommand{\OL}{\Omega_{\Lambda}}

\newcommand{\rmr}{\rho_m}

\newcommand{\rD}{\rho_D}

\newcommand{\rLo}{\rho_{\CC 0}}

\newcommand{\newtext}[1]{{\textcolor{black}{#1}}}

\newcommand{\be}{\begin{equation}}
\newcommand{\ee}{\end{equation}}

\begin{document}



 \hyphenation{nu-cleo-syn-the-sis u-sing si-mu-la-te ma-king
cos-mo-lo-gy know-led-ge e-vi-den-ce stu-dies be-ha-vi-or
res-pec-ti-ve-ly appro-xi-ma-te-ly gra-vi-ty sca-ling
ge-ne-ra-li-zed pa-ra-me-ter ge-ne-ra-ted}




\begin{center}
{\it\LARGE Vacuum dynamics in the Universe versus a rigid $\CC=$const.}\footnote{Invited review paper based on the invited plenary talk by J. Sol\`a  at the Conference on Cosmology, Gravitational Waves and Particles, NTU, Singapore, 2017. Some of the results and discussions presented here have been further updated and slightly expanded with respect to the published version.} \vskip 2mm

 \vskip 8mm

\textbf{Joan Sol\`a, Adri\`a G\'omez-Valent, and Javier de Cruz P\'erez}

\vskip 0.5cm
Departament de F\'isica Qu\`antica i Astrof\'isica, and Institute of Cosmos Sciences,\\ Universitat de Barcelona, \\
Av. Diagonal 647, E-08028 Barcelona, Catalonia, Spain

\vskip0.5cm

\vskip0.4cm

E-mails:   sola@fqa.ub.edu, adriagova@fqa.ub.edu, decruz@fqa.ub.edu

 \vskip2mm

\end{center}
\vskip 15mm

\begin{quotation}
\noindent {\large\it \underline{Abstract}}.
In this year, in which we celebrate 100 years of the cosmological term, $\CC$, in Einstein's gravitational field equations, we are still facing the crucial question whether $\CC$ is truly a fundamental constant or a mildly evolving dynamical variable. After many theoretical attempts to understand the meaning of $\CC$, and in view of the enhanced accuracy of the cosmological observations, it seems now mandatory that this issue should be first settled empirically before further theoretical speculations on its ultimate nature. In this work, we summarize the situation of some of these studies. Devoted analyses made recently show that the $\CC=$const. hypothesis, despite being the simplest, may well not be the most favored one. The overall fit to the cosmological observables SNIa+BAO+$H(z)$+LSS+CMB singles out the class RVM of the ``running'' vacuum models, in which $\CC=\CC(H)$ is an affine power-law function of the Hubble rate. It turns out that the performance of the RVM as compared to the ``concordance'' $\CC$CDM  model (with $\CC=$const.) is much better. The evidence in support of the RVM may reach $\sim 4\sigma$ c.l., and is bolstered with Akaike and Bayesian  criteria providing strong evidence in favor of the RVM option. We also address the implications of this framework on the tension between the CMB and local measurements of the current Hubble parameter.
\end{quotation}
\vskip 5mm

\newpage


\newpage


\section{Introduction}\label{intro}

One hundred years ago, in mid-February 1917, the famous seminal paper in which Einstein introduced the cosmological term $\CC$ (actually denoted ``$\lambda$'' in it), as a part of the generally covariant gravitational field equations, was published\,\cite{Einstein1917}. Fourteen years later, the idea of $\CC$ as a fundamental piece of these equations was abandoned by Einstein himself\,\cite{Einstein1931}; and only one year later the Einstein-de Sitter model, modernly called the cold dark matter model (CDM), was proposed with no further use of the $\CC$ term for the description of the cosmological evolution\,\cite{EinsteindeSitter1932}. The situation with $\CC$ did not stop here and it took an unexpected new turn when $\CC$ reappeared shortly afterwards in the works of Lema\^{\i}tre\,\cite{Lemaitre1934}, wherein $\CC$ was associated to the concept of vacuum energy density through the expression $\rL = \CC/(8\pi G)$ -- in which $G$ is Newton's constant. This association is somehow natural if we take into account that the vacuum energy is thought of as being uniformly distributed in all corners of space, thus preserving the Cosmological Principle. The problem (not addressed by Lema\^{\i}tre) is to understand the origin of the vacuum energy in fundamental physics, namely in the quantum theory and more specifically in quantum field theory (QFT). Here is where the cosmological constant (CC) problem first pops up. It was first formulated in preliminary form  by Zeldovich 50 years ago\,\cite{Zeldovich1967}.

The CC  problem\,\cite{CCP1,CCP2,JSPReview2013} is the main source of headache for every theoretical cosmologist confronting his/her theories with the measured value of $\rL$\,\cite{Planck2015,PlanckDE2015,Planck2016}.  The purported discovery of the Higgs boson at the LHC has actually accentuated the CC problem greatly, certainly much more than is usually recognized\,\cite{JSPReview2013}.
Owing to the necessary spontaneous symmetry breaking (SSB) of the electroweak (EW) theory, there emerges an induced contribution  to $\rL$ that must also be taken into account. These SSB effects are appallingly much larger (viz. $\sim 10^{56}$) than the tiny value $\rL\sim 10^{-47}$ GeV$^4$ extracted from observations\footnote{Being $\rL$ a density, and hence a dimensionful quantity, ``tiny'' value  means only within the particle physics standards, of course.} . So the world-wide celebrated  ``success'' of the Higgs finding  in particle physics actually became a cosmological fiasco, since it automatically detonated the ``modern CC problem'', i.e. the confirmed size of the EW vacuum, which should be deemed  as literally ``real'' (in contrast to other alleged -- ultralarge -- contributions from QFT)  or ``unreal'' as the Higgs boson itself! One cannot exist without the other. Such uncomfortable situation of the Higgs boson with cosmology might be telling us that the found Higgs boson is not a fundamental particle, as in such a case the EW vacuum could not be counted as a fundamental SSB contribution to the vacuum energy of the universe.
We refer the reader to some review papers\,\cite{CCP1,CCP2}, including \cite{JSPReview2013,SolaGomez2015}, for a more detailed presentation of the CC problem.
Setting aside the ``impossible'' task of predicting the $\CC$ value itself -- unless it is understood as a ``primordial renormalization''\,\cite{GRF2015} -- we will focus here on a special class of models in which $\CC$ appears neither as a rigid constant nor as a scalar field (quintessence and the like)\,\cite{CCP2}, but as a ``running'' quantity in QFT in curved spacetime. This is a natural option for an expanding universe. As we will show, such kind of dynamical vacuum models are phenomenologically quite successful; in fact so successful that they are currently challenging the $\CC$CDM\,\cite{JCAP2015b,MNRAS2015,JCAP2015a,BPS2009,Grande2011}, see specially the most recent works\,\cite{ApJL2015,ApJ2017,MPLA2017,PRD2017,PLB2017}, in which the most significant signs of vacuum dynamics have been found. Potential time variation of the fundamental constants associated to these models has also been explored in \cite{FritzschSola2012,Preface, FritzschSola2015, SpecialIssueMPLA,Sola2014,FritzschSolaNunes2017}, and can be used as complementary experimental signatures for them.


\section{Running vacuum as the next-to-minimal step beyond the $\CC$CDM}

 The ``concordance'' or $\CC$CDM model, i.e. the standard model of cosmology is based on the assumption of the existence of dark matter (DM) and the spatial flatness of the Friedmann-Lema\^{\i}tre-Robertson-Walker (FLRW) metric. At the same time the model is based on the existence of a nonvanishing but rigid (and positive) cosmological constant term: $\CC=$const. The model was first proposed as having the minimal ingredients for a possible successful phenomenological description of the data by Peebles in 1984\,\cite{Peebles1984}. Nowadays we know it is consistent with a large body of observations, and in particular with the high precision data from the cosmic microwave background (CMB) anisotropies\,\cite{Planck2015}.

The rigid $\CC=$const. term in the concordance $\CC$CDM model is the simplest (perhaps too simple) possibility. The running vacuum models (RVMs)\,(cf. \cite{JSPReview2013,SolaGomez2015} and references therein) build upon the idea that the cosmological term $\CC$, and the corresponding vacuum energy density, $\rL$,  should be time dependent quantities in cosmology. It is difficult to conceive an expanding universe with a strictly constant vacuum energy density that has remained
immutable since the origin of time. Rather, a smoothly evolving DE density that inherits its time-dependence
from cosmological variables, such as the Hubble rate
$H(t)$, or the scale factor $a(t)$, is not only a qualitatively more
plausible and intuitive idea, but is also suggested by fundamental
physics, in particular by QFT in curved
spacetime. We denote it in general by $\rD=\rD(t)$. Despite its time evolution, it may still have the vacuum equation of state (EoS) $w=-1$, or a more general one $w\neq -1$, or even a dynamical effective EoS $w=w(t)$. The main standpoint of the RVM class of dynamical DE models is that $\rD$ ``runs'' because the effective action receives quantum effects from the matter fields. The leading effects may generically be captured from a renormalization group equation (RGE) of the form\,\cite{JSPReview2013}
\begin{eqnarray}\label{seriesRLH}
\frac{d\rD}{d\ln
\mu^2}=\frac{1}{(4\pi)^2}\sum_{i}\left[\,a_{i}M_{i}^{2}\,\mu^{2}
+\,b_{i}\,\mu^{4}+c_{i}\frac{\mu^{6}}{M_{i}^{2}}\,+...\right] \,.
\end{eqnarray}
The running scale $\mu$ is typically identified with $H$ or related variables. The RVM ansatz is that $\rD=\rD(H)$  because $\mu$ will be naturally associated to the Hubble parameter at a given epoch $H=H(t)$, and hence $\rD$ should evolve with the rate of expansion $H$. Notice  that $\rD(H)$ can involve \emph{only} even powers of
the Hubble rate $H$ (because of the covariance of the effective
action)\,\cite{JSPReview2013}. The coefficients $a_i$, $b_i$, $c_i$... are dimensionless, and the $M_i$ are the masses of the particles in the loops. Because $\mu^2$ can be in general a linear combination of the homogeneous terms $H^2$ and $\dot{H}$, it is obvious that upon integration of the above RGE we expect the following general type of (appropriately normalized) RVM density\,\cite{JSPReview2013,SolaGomez2015,GRF2015}:
\begin{eqnarray}\label{eq:ModelsA}
\rho_D(H)&=&\frac{3}{8\pi
G}\left(C_0+\nu H^2+\tilde{\nu}\dot{H}\right)+{\cal O}(H^4)\,,
\end{eqnarray}
where $\nu$ and $\tilde{\nu}$ are dimensionless parameters, but $C_0$ has dimension $2$ (energy squared) in natural units. We emphasize that $C_0\neq 0$ so as to insure a smooth $\CC$CDM limit when the dimensionless coefficients $\nu$ and $\tilde{\nu}$ are set to zero\,\footnote{It is important to make clear that models with $C_0=0$ (for any $\nu$ and $\tilde{\nu}$) are ruled out by the observations, as shown in \cite{JCAP2015b,MNRAS2015,JCAP2015a}. This conclusion also applies to all DE models of the form  $\rD\sim aH+bH^2$,
with a linear term $\sim H$ admitted only on phenomenological grounds\,\cite{JCAP2015b,MNRAS2015}. In particular, the model $\rD\sim H$ is strongly ruled out, see \cite{MNRAS2015} (and the discussion in its Appendix).}. The interesting possibility that $\nu$ and/or $\tilde{\nu}$ are nonvanishing may induce a time evolution of the vacuum energy density. These dimensionless coefficients can be computed in QFT from the ratios squared of the masses to the Planck mass\,\cite{Fossil07}, and are therefore small as expected from their interpretation as $\beta$-function coefficients of the RGE (\ref{seriesRLH}). Since some of the masses inhabit the GUT scale $M_X\sim 10^{16}$ GeV, the values of $\nu,\tilde{\nu}$ need not be very small, typically $\sim 10^{-3}$ at most upon accounting for the large multiplicities that are typical in a GUT -- see Ref.\cite{Fossil07} for a concrete estimate.

Finally, we note that the ${\cal O}(H^4)$-terms in (\ref{eq:ModelsA}) are irrelevant for the study of the current universe, but are essential for the correct account of the inflationary epoch in this context and to explain the graceful exit and entropy problems\,\cite{SolaGomez2015,GRF2015}. The RVM (\ref{eq:ModelsA}) is therefore capable of providing a unified dynamical vacuum picture for the entire cosmic evolution\,\cite{LimBasSol2013,LimBasSol2015,LimBasSol2016}.

Concerning the parameters $\nu$ and $\tilde{\nu}$, they must be determined phenomenologically by confronting the model to the wealth of observations. It is remarkable that the aforementioned theoretical estimate\,\cite{Fossil07}, leading to $\nu,\tilde{\nu}\sim 10^{-3}$, is of the order of magnitude of the phenomenological determination\,\cite{ApJL2015,JCAP2015b,MNRAS2015,JCAP2015a,BPS2009,Grande2011}. For the current presentation, however, we will assume $\tilde{\nu}=0$ hereafter and will focus on the implications for the current universe of the canonical RVM:
\begin{eqnarray}\label{eq:CanonicalRVM}
\rho_D(H)&=&\frac{3}{8\pi
G}\left(C_0+\nu H^2\right)\,.
\end{eqnarray}
We will use the above expression to study the RVM background cosmology and the corresponding perturbations equations. We can compare with the concordance $\CC$CDM cosmology by just setting $\nu=0$ in the obtained results. The background cosmological equations for the RVM take on the same form as in the $\CC$CDM by simply replacing the rigid cosmological term with the dynamical vacuum energy density (\ref{eq:CanonicalRVM}). In this way we obtain the generalized Friedmann's and acceleration equations, which read as follows:
 \begin{equation}\label{eq:generalizedFriedmann}
3H^2=8\pi
G(\rho_m+\rD(H))\,,\ \ \ \ \ \ \ \  2\dot{H}+3H^2=-8\pi
G(w_m\rho_m+w_D\rD(H))\,,
\end{equation}
where  $w_m$ and $w_D$ are the  EoS parameters of the matter fluid and of the DE, respectively. It is well known that $w_m=1/3,0$ for relativistic and nonrelativistic matter, respectively. Then, for simplicity, we will denote the EoS of the DE component simply as $w$. The explicit solution will depend of course on whether we assume that the DE is canonical vacuum energy ($w=-1$), in which case $\rD$ can be properly denoted as $\rL$, or dynamical DE with $w\neq -1$. In some cases $w$ can also be a function of time or of some cosmic variable, but we shall not consider this possibility here -- see, however, Ref.\,\cite{JCAP2015b} in which such situation would be mandatory. The solution of the cosmological equations may also depend on whether the gravitational coupling $G$ is constant or also running  with the expansion, $G=G(H)$ (as $\rD$ itself). And, finally, it may depend on whether we assume that there exists an interaction of the DE with the matter (mainly DM). Whatever it be the  nature of our assumptions on these important details, they must be of course consistent with the Bianchi identity, which is tantamount to saying with the local covariant conservation laws. In fact, these possibilities have all been carefully studied in the literature and the complete solution of the cosmological equations has been provided in each case\,\cite{ApJ2017,MPLA2017,PLB2017,ApJL2015,PRD2017,JCAP2015b,MNRAS2015,JCAP2015a,BPS2009,Grande2011}.  In what follows we report only on some of the solutions for the densities  of matter and DE in the case when there is an interaction between the DE and matter at fixed $G$. At the same time we will compare the result when the EoS of the DE is $w\neq -1$ (which means that we will leave this parameter also free in the fit) and when $w=-1$ (the strict vacuum case).


\section{Canonical RVM with
conserved baryon and radiation densities}

The total matter density $\rho_m$ can be split into the contribution from baryons and cold DM, namely $\rho_m=\rho_b+\rho_{dm}$. In the following we assume that the  DM density is the only one that carries the anomaly, whereas radiation and baryons are self-conserved, so that their energy densities evolve in the standard way $\rho_r(a)=\rho_{r}^0\,a^{-4}$ and $\rho_b(a) = \rho_{b}^0\,a^{-3}$. On the other hand the dynamical evolution of the vacuum is given by Eq.\,\eqref{eq:CanonicalRVM}.  Since it is only the DM that exchanges energy with the vacuum, the local conservation law reads as follows:
\begin{equation}\label{eq:Qequations}
\dot{\rho}_{dm}+3H\rho_{dm}=Q\,,\ \ \ \ \ \ \, \dot\rho_{\CC}=-{Q}\,.
\end{equation}
The source function $Q$ is a calculable expression from \eqref{eq:CanonicalRVM} and Friedmann's equation in \eqref{eq:generalizedFriedmann}. We find:
$Q=-\dot{\rho}_{\Lambda}=\nu\,H(3\rho_{dm}+3\rho_{b}+4\rho_r)=\nu\,H(3\rho_{m}+4\rho_r)$.
It can be useful to compare the canonical RVM with two alternative dynamical vacuum models (DVMs) with different forms of the interaction sources. Let us therefore list the three models under comparison, which we denote  RVM, $Q_{dm}$ and $Q_{\CC}$ according to the structure of the interaction source, or also for convenience just I, II and III:
\begin{eqnarray}\label{eq:QforModelRVM}
&&{\rm Model\ I\ \ }(w{\rm RVM}):\ Q=\nu\,H(3\rho_{m}+4\rho_r)\label{eq:QforModelQdm}\\
&&{\rm Model\ II\ \ }(wQ_{dm}):\ Q_{dm}=3\nu_{dm}H\rho_{dm}\\
&&{\rm Model\ III\ \ }(wQ_{\CC}):\ Q_{\CC}=3\nu_{\CC}H\rho_{\CC}\,.\label{eq:QforModelQL}
\end{eqnarray}
Each model has a characteristic (dimensionless) parameter $\nu_i=\nu,\nu_{dm},\nu_{\CC}$ as a part of the interaction source, which must be fitted to the observational data. Notice that the three model names are preceded by $w$  to recall that, in the general case, the equation of state (EoS) is very near to the vacuum one ($w=-1+\epsilon$, with $|\epsilon|\ll1$). For this reason these dynamical quasi-vacuum models  are also denoted as $w$DVMs. In the particular case $w=-1$ (i.e. $\epsilon=0$) the $w$DVMs become just the canonical DVMs.
As an example, let us provide the matter and vacuum energy densities for the RVM:
\begin{eqnarray}\label{eq:rhomRVMc}
\rho_{dm}(a) &=& \rho_{dm}^0\,a^{-3(1-\nu)} + \rho_{b}^0\left(a^{-3(1-\nu)} - a^{-3}\right)- \frac{4\nu\rho^{0}_r}{1+3\nu}\left(a^{-4}-a^{-3(1-\nu)}\right)
\end{eqnarray}
and
\begin{eqnarray}\label{eq:rLRVMc}
\rho_\Lambda(a) &=& \rLo + \frac{\nu\,\rho_{m}^0}{1-\nu}\left(a^{-3(1-\nu)}-1\right)\nonumber\\
 &+& \frac{\nu}{1-\nu}\,\,\rho^{0}_{r}\left(\frac{1-\nu}{1+3\nu}\,a^{-4} + \frac{4\nu}{1+3\nu}\,a^{-3(1-\nu)}-1\right).
\end{eqnarray}
For the corresponding expressions of the other models, see\,\cite{PRD2017,PLB2017}. Models II and III were previously studied in different approximations e.g. in \cite{Salvatelli2014,Murgia2016,Li2016}.
As can be easily checked, for $\nu_i\to 0$ we recover the corresponding results for the $\CC$CDM, as it should.
The Hubble function can be immediately obtained from these formulas after inserting them in Friedmann's equation, together with the conservation laws for radiation and baryons,  $\rho_r(a)=\rho_{r}^0\,a^{-4}$ and $\rho_b(a) = \rho_{b}^0\,a^{-3}$.

\section{The XCDM and CPL parametrizations}
In  Section \ref{sect: Fitting}, when we compare the DVMs to the $\CC$CDM, it is also convenient to fit the data  to the simple XCDM parametrization of the dynamical DE\,\cite{XCDM}. Since both matter and DE are self-conserved (i.e., they are not interacting) in the XCDM, the DE energy density as a function of the scale factor is simply given by $\rho_X(a)=\rho_{X0}\,a^{-3(1+w_0)}$, with $\rho_{X0}=\rho_{\CC 0}$, where $w_0$ is the (constant) equation of state (EoS) parameter of  the generic DE entity $X$ in this parametrization. The normalized Hubble function is:
\begin{equation}\label{eq:HXCDM}
E^2(a)=\Omega_m\,a^{-3}+\Omega_r\,a^{-4}+\OL\,a^{-3(1+w_0)}\,.
\end{equation}
For $w_0=-1$ it boils down to that of the $\CC$CDM with rigid CC term. Use of the XCDM parametrization becomes useful so as to roughly mimic a (noninteractive) DE scalar field with constant EoS. For $w_0\gtrsim-1$ the XCDM mimics quintessence, whereas for $w_0\lesssim-1$ it mimics phantom DE.

A slightly more sophisticated parametrization to the behavior of a noninteractive scalar field playing the role of dynamical DE is furnished by the CPL parametrization\,\cite{CPL}, in which one assumes that the generic DE entity $X$ has a slowly varying EoS of the form
\begin{equation}\label{eq:CPL}
w_D=w_0+w_1\,(1-a)=w_0+w_1\,\frac{z}{1+z}\,.
\end{equation}
The CPL parametrization, in contrast to the XCDM one, gives room for a time evolution of the dark energy EoS owing to the presence of the additional parameter $w_1$, which satisfies $0<|w_1|\ll|w_0|$, with $w_0\gtrsim -1$ or $w_0\lesssim -1$.
The corresponding normalized Hubble function for the CPL can be easily computed:
\begin{equation}
 E^2(a) = \Omega_m\,a^{-3}+ \Omega_r a^{-4}+\OL
 a^{-3(1+w_0+w_1)}\,e^{-3\,w_1\,(1-a)}\,.
\label{eq:HCPL}
\end{equation}
Both the XCDM and the CPL parametrizations can be thought of as a kind of baseline frameworks to be referred to in the study of dynamical DE. They can be used as fiducial models to which we can compare other, more sophisticated, models for the dynamical DE, such as the DVMs under study.
The XCDM, however, is more appropriate for a fairer comparison with the DVMs, since they have one single vacuum parameter $\nu_i$. For this reason we present the main fitting results with the XCDM, along with the other models, in Table 1. The numerical fitting results for the CPL parametrization are given in \cite{PRD2017}. Owing to the presence of an extra parameter the errors in the fitting values of $w_0$ and $w_1$ are bigger than the error in the single parameter $w_0$ of the XCDM parametrization. For this presentation, we limit ourselves to report on the latter, together with the rest of the models.

\section{Structure formation under vacuum dynamics}

The DVMs and $w$DVMs are characterized by a dynamical vacuum/quasi-vacuum energy density. Therefore, in order to correctly fit these models to the data, a generalized treatment of the linear structure formation beyond the $\CC$CDM is of course mandatory. At the (subhorizon) scales under consideration we will neglect the perturbations of the vacuum energy density in front of the perturbations of the matter field. This has been verified for various related cases, see e.g. \cite{GrandePelinsonSola2009} and \cite{JCAP2015b} . For a recent detailed study directly involving the DVMs under consideration, see\,\cite{GomSol2017}.
 In the presence of dynamical vacuum energy the matter density contrast $\delta_m=\delta\rho_m/\rho_m$ obeys the following differential equation (cf. \,\cite{PRD2017,PLB2017} for details):
\begin{equation}\label{diffeqD}
\delta^{\prime\prime}_m(a) + \frac{A(a)}{a}\delta_{m}^{\prime}(a) + \frac{B(a)}{a^2}\delta_m(a) = 0  \,,
\end{equation}
where the prime denotes differentiation with respect to the scale factor.
The functions $A(a)$ and $B(a)$ can be determined after a straightforward application of the general perturbation
equations in the presence of dynamical vacuum (cf. \cite{GrandePelinsonSola2009}), with the result:
\begin{eqnarray}
&&A(a) = 3 + \frac{aH^{'}}{H} + \frac{\Psi}{H} - 3r\epsilon\label{eq:Afunction}\\
&&B(a) = -\frac{4\pi{G}\rho_m }{H^2} + 2\frac{\Psi}{H} + \frac{a\Psi^{'}}{H} -15r\epsilon
 - 9\epsilon^{2}r^{2} +3\epsilon(1+r)\frac{\Psi}{H} -3r\epsilon\frac{aH^{'}}{H}\,. \label{eq:Bfunction}
\end{eqnarray}
Here $r \equiv \rho_\Lambda/\rho_m$ and  $\Psi\equiv Q/{\rmr}$. {For $\nu_i=0$ we have $\Psi=3Hr\epsilon$}, and after some calculations one can easily show that (\ref{diffeqD})  can be brought back to the common form for the XCDM and $\CC$CDM.
The (vacuum-matter) interaction source $Q$ for each DVM is given by \eqref{eq:QforModelRVM}-\eqref{eq:QforModelQL}. For $\rL=$const. and for the XCDM  there is no such an interaction, therefore $Q=0$, and Eq.\,(\ref{diffeqD}) reduces to the $\CC$CDM form:
\begin{equation}\label{homog}
\delta_m''(a)+\left(\frac{3}{a}+\frac{H'(a)}{H(a)}\right)\,\delta_m'(a)-\frac{4\pi{G}\rho_m(a) }{H^2(a)}\,\frac{\delta_m(a)}{a^2}=0\,.
\end{equation}
Recalling that $\rho_m(a)=\rho_m^0\,a^{-3}$ in the $\CC$CDM, the growing mode solution of this equation can be solved by
quadrature:
\begin{equation}\label{Da}
\delta_m(a)=\frac{5\Omega_m}{2}
\frac{H(a)}{H_0}\int^a_0\frac{d\tilde{a}}{(\tilde{a}\,H(\tilde{a})/H_0)^3}\,.
\end{equation}
One can easily check that in the matter-dominated epoch ($\Omega_m(a)\simeq1$, $H^2\simeq H_0^2  \Omega_m a^{-3}$), the above equation yields $\delta_m\simeq a$, as expected. However, being  $\Omega_m(a)<1$  there is an effective suppression of the form $\delta_m\simeq a^s$ with $s<1$\,\cite{Grande2007}. The last feature is also true in the presence of dynamical vacuum. However, the generalized perturbations equation \eqref{diffeqD} cannot be solved analytically and one has to proceed numerically.  The first thing to do is to fix the initial conditions analytically. This is possible because  at high redshift, namely when nonrelativistic matter dominates over radiation and DE, functions (\ref{eq:Afunction}) and (\ref{eq:Bfunction}) are then approximately constant and Eq.\,(\ref{diffeqD}) admits power-law solutions $\delta_m(a) = a^{s}$. The values for the power $s$ can be computed for each model. For example, for the $w$RVM it can be shown that $s=1 + ({3}/{5})\nu\left(\frac{1}{w} -4\right) + \mathcal{O}(\nu^2)$. Notice that $s<1$ for $\nu>0$ and $w$ near $-1$. Using the appropriate initial conditions for each model, the numerical solution of (\ref{diffeqD}) can be obtained -- see \cite{PRD2017,PLB2017} for more details.

Armed with these equations, the linear LSS regime is usually analyzed with the help of the weighted linear growth $f(z)\sigma_8(z)$, where $f(z)=d\ln{\delta_m}/d\ln{a}$ is the growth rate and $\sigma_8(z)$ is the rms mass fluctuation on $R_8=8\,h^{-1}$ Mpc scales. It is computed as follows (see e.g. \cite{PRD2017}):
\begin{equation}
\begin{small}\sigma_{\rm 8}(z)=\sigma_{8, \CC}
\frac{\delta_m(z)}{\delta^{\CC}_{m}(0)}
\sqrt{\frac{\int_{0}^{\infty} k^{n_s+2} T^{2}(\vec{p},k)
W^2(kR_{8}) dk} {\int_{0}^{\infty} k^{n_{s,\CC}+2} T^{2}(\vec{p}_\Lambda,k) W^2(kR_{8,\Lambda}) dk}}\,,\label{s88general}
\end{small}\end{equation}
where $W$ is a top-hat smoothing function and $T(\vec{p},k)$ the transfer function\,\cite{PRD2017}. The fitting parameters for each model are contained in $\vec{p}$.
Following the above mentioned references, we define as fiducial model the $\CC$CDM at fixed parameter values from the Planck 2015 TT,TE,EE+lowP+lensing data\,\cite{Planck2015}. These fiducial values are collected in  $\vec{p}_\CC$.  The theoretical calculation of $\sigma_{\rm 8}(z)$ and of the product $f(z)\sigma_8(z)$ for each model is essential to compare with the LSS formation data. The calculation is possible after obtaining the fitting results for the parameters, as indicated in Tables 1 and 2 and Fig. 1. The result for  $f(z)\sigma_8(z)$ is plotted in Fig. 2, together with the observational points. In Sect.\ref{sec:RunMass} we will further discuss these results. In the next section we discuss some basic facts of the fit analysis.

\begin{table*}
\begin{center}
\resizebox{1\textwidth}{!}{
\begin{tabular}{ |c|c|c|c|c|c|c|c|c|c|}
\multicolumn{1}{c}{Model} &  \multicolumn{1}{c}{$h$} &  \multicolumn{1}{c}{$\omega_b$} & \multicolumn{1}{c}{{\small$n_s$}}  &  \multicolumn{1}{c}{$\Omega_m$} &\multicolumn{1}{c}{$\nu_i$} &\multicolumn{1}{c}{$w$} &\multicolumn{1}{c}{$\chi^2_{\rm min}/dof$} & \multicolumn{1}{c}{$\Delta{\rm AIC}$} & \multicolumn{1}{c}{$\Delta{\rm BIC}$}\vspace{0.5mm}
\\\hline
$\Lambda$CDM  & $0.688\pm 0.004$ & $0.02243\pm 0.00013$ &$0.973\pm 0.004$& $0.298\pm 0.004$ & - & -  & 84.40/85 & - & - \\
\hline
XCDM  & $0.672\pm 0.006$& $0.02251\pm0.00013 $&$0.975\pm0.004$& $0.311\pm0.006$ & - &$-0.936\pm{0.023}$  & 76.80/84 & 5.35 & 3.11 \\
\hline
RVM  & $0.674\pm 0.005$& $0.02224\pm0.00014 $&$0.964\pm0.004$& $0.304\pm0.005$ &$0.00158\pm 0.00042 $ & -  & 68.67/84 & 13.48 & 11.24 \\
\hline
$Q_{dm}$  & $0.675\pm 0.005$& $0.02222\pm0.00014 $&$0.964\pm0.004$& $0.304\pm0.005$ &$0.00218\pm 0.00057 $&-  & 69.13/84 & 13.02 &10.78 \\
\hline
$Q_\Lambda$  & $0.688\pm 0.003$& $0.02220\pm0.00015 $&$0.964\pm0.005$& $0.299\pm0.004$ &$0.00673\pm 0.00236 $& -  &  76.30/84 & 5.85 & 3.61\\
\hline
$w$RVM  & $0.671\pm 0.007$& $0.02228\pm0.00016 $&$0.966\pm0.005$& $0.307\pm0.007$ &$0.00140\pm 0.00048 $ & $-0.979\pm0.028$ & 68.15/83 & 11.70 & 7.27 \\
\hline
$w{Q_{dm}}$  & $0.670\pm 0.007$& $0.02228\pm0.00016 $&$0.966\pm0.005$& $0.308\pm0.007$ &$0.00189\pm 0.00066 $& $-0.973\pm 0.027$ & 68.22/83 & 11.63 & 7.20\\
\hline
$w{Q_\Lambda}$  & $0.671\pm {0.007}$& $0.02227\pm0.00016 $&$0.965\pm0.005$& $0.313\pm0.006$ &$0.00708\pm 0.00241 $& $-0.933\pm0.022$ &   68.24/83 & 11.61 & 7.18\\
\hline
\end{tabular}}
 \caption{
{\scriptsize Best-fit values for the free parameters of the $\CC$CDM,  XCDM, the three dynamical vacuum models (DVMs) and the three dynamical quasi-vacuum models ($w$DVMs), including their statistical significance (i.e. the values of the $\chi^2$-test and the difference of the values of the Akaike and Bayesian information criteria, AIC and BIC, with respect to the $\CC$CDM).
For a detailed description of the data and a full list of observational references, see \cite{ApJ2017,PRD2017}. The quoted number of degrees of freedom ($dof$) is equal to the number of data points minus the number of independent fitting parameters ($4$ for the $\CC$CDM, 5 for the XCDM and the DVMs, and 6 for the $w$DVMs). For the CMB data we have used the marginalized mean values and {covariance matrix} for the parameters of the compressed likelihood for Planck 2015 TT,TE,EE + lowP+ lensing as in \cite{PLB2017}. Each best-fit value and the associated uncertainties have been obtained by marginalizing over the remaining parameters.}}
 \label{Table1}
\end{center}
\end{table*}

\section{Fitting the DVMs to observations}\label{sect: Fitting}

In this section, we put the dynamical vacuum models (DVMs) discussed above to the test, see \,\cite{ApJL2015,ApJ2017,MPLA2017,PRD2017,PLB2017} for more details.  It proves useful to study them altogether in a comparative way, and of course we compare them to the $\CC$CDM.

We confront all these models to the main set of cosmological observations compiled to date, namely we fit the models to the following wealth of data (cf. Ref.\,\cite{ApJ2017,MPLA2017,PRD2017} for a complete list of references): i) the data from distant type Ia supernovae (SNIa); ii) the data on baryonic acoustic oscillations (BAO's); iii) the known values of the Hubble parameter at different redshift points, $H(z_i)$; iv) the large scale structure (LSS) formation data encoded in the weighted linear growth rate $f(z_i)\sigma_8(z_i)$;  v) the CMB distance priors from Planck 2015.  Thus, we use the essential observational data represented by the cosmological observables SNIa+BAO+$H(z)$+LSS+CMB.  For the analysis we have defined a joint likelihood function ${\cal L}$ from the product of the likelihoods for all the data sets discussed above. For Gaussian errors, the total $\chi^2$ to be minimized reads:
\be
\chi^2_{tot}=\chi^2_{SNIa}+\chi^2_{BAO}+\chi^2_{H}+\chi^2_{f\sigma_8}+\chi^2_{CMB}\,.
\ee
Each one of these terms is defined in the standard way,
including the covariance matrices for each sector\,\footnote{See the details in the appendix of \cite{PRD2017}.}.
The fitting results for models \eqref{eq:QforModelRVM}-\eqref{eq:QforModelQL} are displayed in terms of contour plots in Fig.\,1.



\begin{figure}[t]
\centering
\includegraphics[angle=0,width=1.03\linewidth]{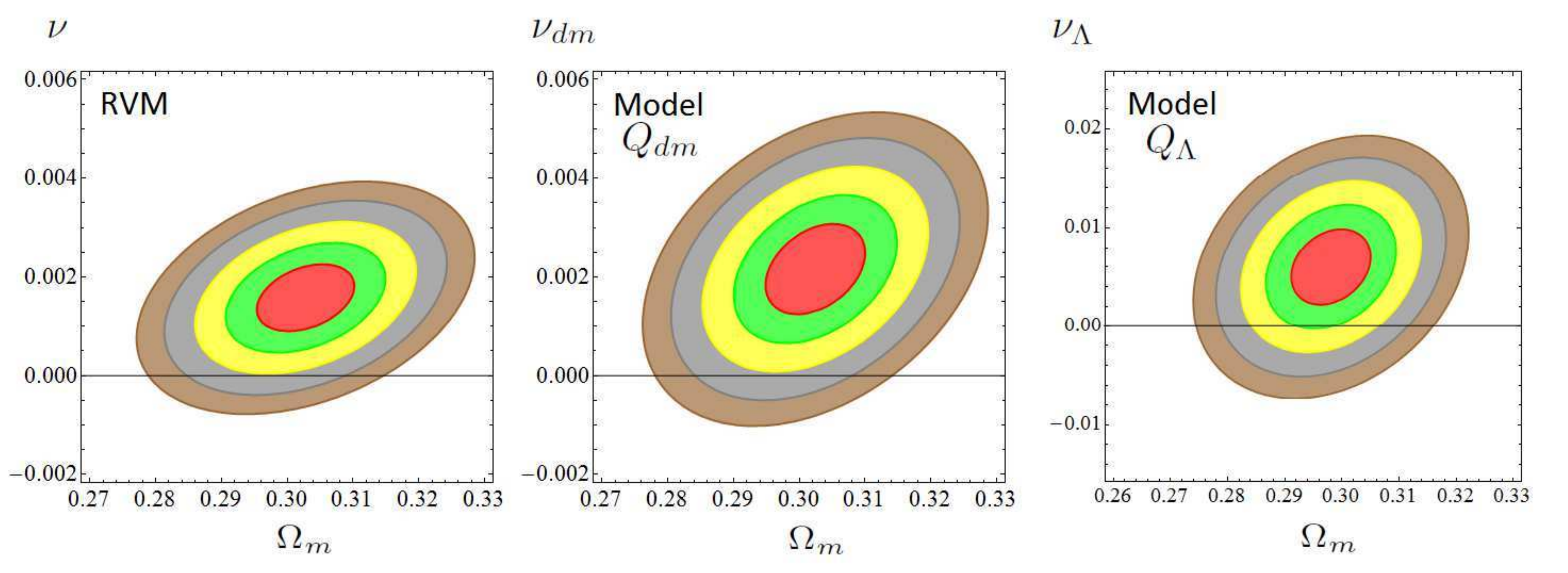}
\caption{\label{fig:CLRVMandQdm}%
\scriptsize {Likelihood contours in the $(\Omega_m,\nu_i)$-plane for the three DVMs I, II and III  (we are restricting here to the vacuum case $w=-1$ in all cases) defined in \eqref{eq:QforModelRVM}-\eqref{eq:QforModelQL}. Shown are the regions corresponding to $-2\ln\mathcal{L}/\mathcal{L}_{max}=2.30$, $6.18, 11.81$, $19.33$, $27.65$ (corresponding to 1$\sigma$, 2$\sigma$, 3$\sigma$, 4$\sigma$ and 5$\sigma$ c.l.) after marginalizing over the rest of the fitting parameters indicated in Table 1. The elliptical shapes have been obtained applying the standard Fisher approach. We estimate that for the RVM, $94.80\%$ (resp. $89.16\%$) of the 4$\sigma$ (resp. 5$\sigma$) area is in the $\nu>0$ region. The $\CC$CDM ($\nu_i=0$) appears disfavored at $\sim 4\sigma$ c.l. in the RVM and $Q_{dm}$, and at $\sim 2.5\sigma$ c.l. for $Q_\CC$. For more details, see \cite{PRD2017}.}
}
\end{figure}
%


For the numerical fitting details, cf. Table 1.
In such table we assess also the comparison of the various models in terms of the time-honored Akaike and Bayesian information criteria, AIC and BIC\,\cite{Akaike,Schwarz}.
These information criteria are extremely useful for comparing different models in competition. The reason is obvious: the models having more parameters have a larger capability to adjust the observations, and for this reason they should be penalized accordingly. It means that the minimum value of  $\chi^2$ should be appropriately corrected so as to take into account this feature. This is achieved through the AIC and BIC estimators, which can be thought of as a modern quantitative formulation of Occam's razor.  They are defined as follows\,\cite{Akaike,Schwarz,KassRaftery1995,Burnham}:
\be
{\rm AIC}=\chi^2_{\rm min}+\frac{2nN}{N-n-1}
\ee
and
\be
{\rm BIC}=\chi^2_{\rm min}+n\,\ln N\,.
\ee
Here $n$ is the number of independent fitting parameters and $N$ the number of data points.
The larger are the differences $\Delta$AIC ($\Delta$BIC) with respect to the model that carries smaller value of AIC (BIC) -- the DVMs here -- the higher is the evidence against the model with larger value of  AIC (BIC) -- the $\CC$CDM.
For $\Delta$AIC and/or $\Delta$BIC in the range $6-10$ we can speak of ``strong evidence'' against the $\CC$CDM, and hence in favor of the DVMs. Above 10, we  are entitled to claim ``very strong evidence''\,\cite{Akaike,Schwarz,KassRaftery1995,Burnham} in favor of the DVMs.

From Table 1 we can read off the results: for Models I and II we find $\Delta$AIC$>10$ {\it and} $\Delta$BIC$>10$ simultaneously.  The results from both are consistent with the fact that these models yield a nonvanishing value for $\nu_i$ with the largest significance ($\sim 4\sigma$). It means that the DVMs I and II fit better the overall data than the $\CC$CDM at such confidence level. Model III (i.e. $Q_{\CC}$) also fits better the observations than the $\CC$CDM, but with a lesser c.l. Indeed, in this case $\Delta$AIC$>5$ {\it and} $\Delta$BIC$>3$, and the parameter $\nu_{\CC}$ is nonvanishing at near $3\sigma$. Thus, the three DVMs are definitely more favored than the $\CC$CDM, and the most conspicuous one is the RVM, Eq.\, (\ref{eq:CanonicalRVM}).

We conclude that the wealth of cosmological data at our disposal currently suggests that the hypothesis $\CC=$const. despite being the simplest may well not be the most favored one. The absence of vacuum dynamics is excluded at nearly $4\sigma$ c.l. as compared to the best DVM considered here (the RVM). The strength of this statement is riveted with the firm verdict of Akaike and Bayesian criteria. Overall we have collected a fairly strong statistical support of the conclusion that the SNIa+BAO+$H(z)$+LSS+CMB cosmological data do favor a mild dynamical vacuum evolution.

\begin{table*}
\begin{center}
\resizebox{1\textwidth}{!}{
\begin{tabular}{ |c|c|c|c|c|c|c|c|c|c|}
\multicolumn{1}{c}{Model} &  \multicolumn{1}{c}{$h$} &  \multicolumn{1}{c}{$\omega_b$} & \multicolumn{1}{c}{{\small$n_s$}}  &  \multicolumn{1}{c}{$\Omega_m$} &\multicolumn{1}{c}{$\nu_i$} &\multicolumn{1}{c}{$w$} &\multicolumn{1}{c}{$\chi^2_{\rm min}/dof$} & \multicolumn{1}{c}{$\Delta{\rm AIC}$} & \multicolumn{1}{c}{$\Delta{\rm BIC}$}\vspace{0.5mm}
\\\hline
$\Lambda$CDM  & $0.690\pm 0.003$ & $0.02247\pm 0.00013$ &$0.974\pm 0.003$& $0.296\pm 0.004$ & - & -  & 90.59/86 & - & - \\
\hline
XCDM  & $0.680\pm 0.006$& $0.02252\pm0.00013 $&$0.975\pm0.004$& $0.304\pm0.006$ & - &$-0.960\pm{0.023}$  & 87.38/85 & 0.97 & -1.29 \\
\hline
RVM  & $0.679\pm 0.005$& $0.02232\pm0.00014 $&$0.967\pm0.004$& $0.300\pm0.004$ &$0.00133\pm 0.00040 $ & -  & 78.96/85 & 9.39 & 7.13 \\
\hline
$Q_{dm}$  & $0.679\pm 0.005$& $0.02230\pm0.00014 $&$0.966\pm0.004$& $0.300\pm0.004$ &$0.00185\pm 0.00057 $&-  & 79.17/85 & 9.18 & 6.92 \\
\hline
$Q_\Lambda$  & $0.690\pm 0.003$& ${0.02224}\pm0.00016 $&$0.965\pm0.005$& $0.297\pm0.004$ &$0.00669\pm 0.00234 $& -  &  82.48/85 & 5.87 & 3.61\\
\hline
$w$RVM  & $0.680\pm 0.007$& $0.02230\pm0.00015 $&$0.966\pm0.005$& $0.300\pm0.006$ &$0.00138\pm 0.00048 $ & $-1.005\pm0.028$ & 78.93/84 & 7.11 & 2.66 \\
\hline
$w{Q_{dm}}$  & $0.679\pm 0.007$& $0.02230\pm0.00016 $&$0.966\pm0.005$& $0.300\pm0.006$ &$0.00184\pm 0.00066 $& $-0.999\pm 0.028$ & 79.17/84 & 6.88 & 2.42\\
\hline
$w{Q_\Lambda}$  & $0.679\pm 0.006$& $0.02227\pm0.00016 $&$0.966\pm0.005$& $0.306\pm0.006$ &$0.00689\pm 0.00237 $& $-0.958\pm0.022$ &   78.98/84 & 7.07 & 2.61\\
\hline
\end{tabular}}
 \caption{
{\scriptsize Best-fit values for the free parameters of the  $\CC$CDM, XCDM, the three dynamical vacuum models (DVMs) and the three dynamical quasi-vacuum models ($w$DVMs), including their statistical significance (the values of the $\chi^2$-test and the difference of the Akaike and Bayesian information criteria AIC and BIC with respect to the $\CC$CDM), i.e. the same as in Table 1 but now adding the $H_0^{\rm Riess}$ local measurement, as indicated in Eq.\,(\ref{eq:H0Riess}).}}
\label{Table1}
\end{center}
\end{table*}



\section{Dynamical vacuum and the $H_0$ tension. A case study.}\label{sec:RunMass}

The framework outlined in the previous sections suggest that owing to a possible small interaction with matter, the vacuum energy density might be evolving with the cosmic expansion. This opens new vistas for a possible understanding of the well known discrepancy between the CMB measurements of $H_0$\,\cite{Planck2015,Planck2016}, suggesting a value below $70$ km/s/Mpc, and the local determinations emphasizing a higher range clearly above $70$ km/s/Mpc\,\cite{RiessH02016}.

\subsection{A little of history}
Such tension is reminiscent of the prediction by the famous astronomer A. Sandage in the sixties, who asserted\,\cite{Sandage1961} that the main task of future observational cosmology would be the search for two parameters: the Hubble constant $H_0$ and the deceleration parameter $q_0$. The first of these parameters defines the most important distance (and time) scale in cosmology prior to any other cosmological quantity. Sandage's last published value with Tammann (in 2010) is $62.3$ km/s/Mpc\,\cite{TammannSandage2010} -- a value that was slightly revised in Ref.\,\cite{TammannReindl2013} as $H_0 = 64.1 \pm
2.4$ km/s/Mpc after due account of the high-weight TRGB (tip
of the red-giant branch) calibration of SNIa.

\begin{figure}[t]
\centering
\includegraphics[angle=0,width=0.6\linewidth]{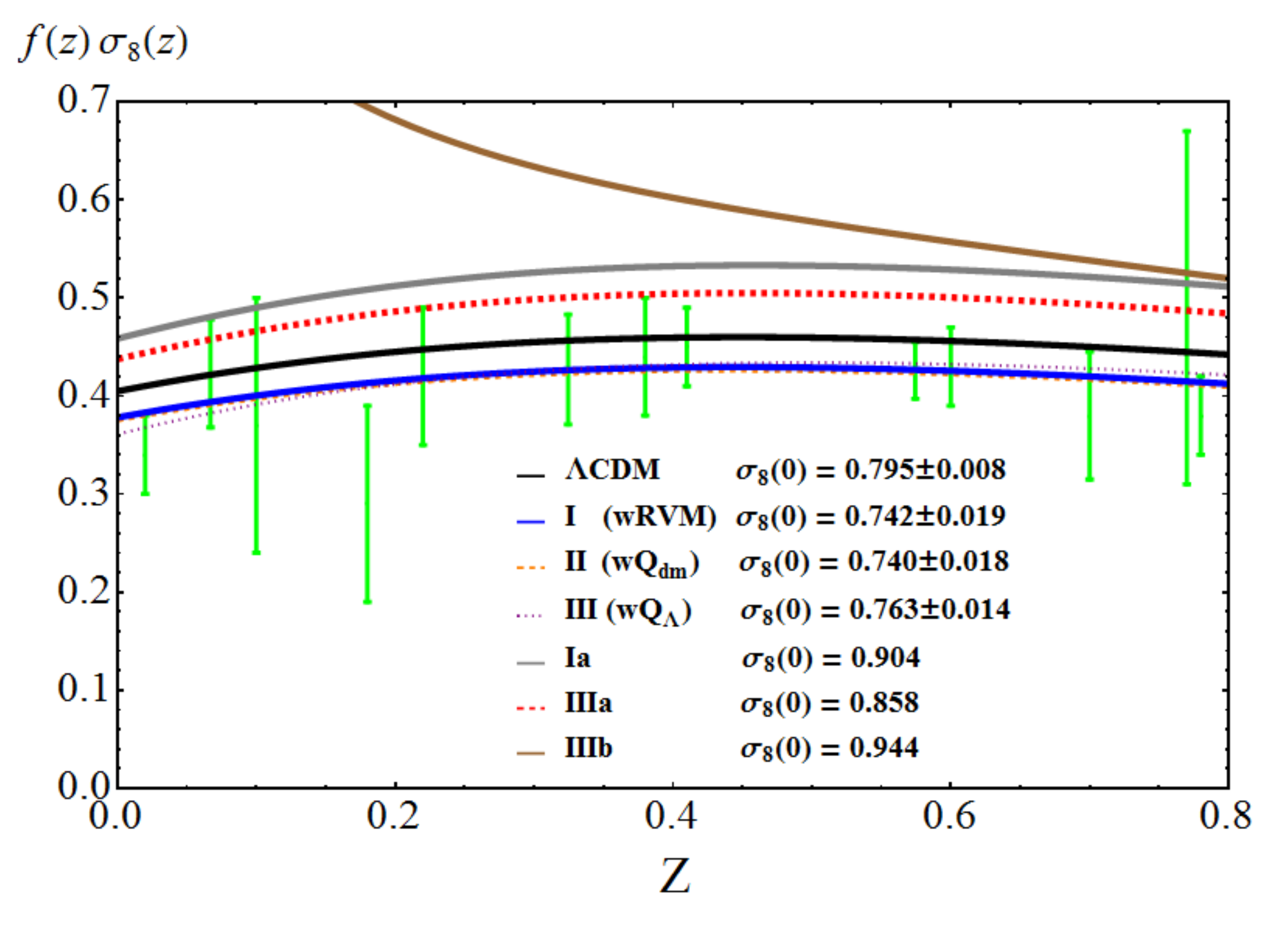}
\caption{\label{fig:CLQv}%
\scriptsize{The large scale structure (LSS) formation data ($f(z)\sigma_8(z)$) and the theoretical predictions for models I, II and III in the case $w\neq- 1$ (i.e. the $w$DVMs). The computed values of $\sigma_8(0)$ for each model are also indicated. The curves for the cases Ia, IIIa and IIIb correspond to special scenarios for Models I and III, in which only the BAO and CMB data are used (not the LSS). Despite the agreement of the CMB measurement $H_0^{\rm Planck}$ with the  Riess et al. local value $H_0^{\rm Riess}$ can be better for these special scenarios, the price to enforce such ``agreement'' is that the concordance with the LSS data is now spoiled (the curves for Ia and IIIa are higher). Case IIIb is our theoretical calculation of the impact on the LSS data for the scenario proposed in \cite{Melchiorri2017b}, aimed at optimally relaxing the tension with $H_0^{\rm Riess}$. Unfortunately it is in severe disagreement with the LSS data. The last three scenarios lead to phantom-like DE and are all disfavored (at different degrees) by the LSS data\,\cite{PLB2017}. }
}
\end{figure}
%

As for the deceleration parameter, $q_0$, its measurement is equivalent to determining $\CC$ in the context of the concordance model.
As indicated in the introduction, on fundamental grounds understanding the value of $\CC$ may not just be a matter of observation; in actual fact it embodies one of the most important and unsolved conundrums of theoretical physics and cosmology of all times: the cosmological constant problem\,\cite{CCP1}\footnote{There is a famous saying by Allan Sandage: `...it is not a matter of debate, it is a matter of observation'\,\cite{TammannReindl2013}, which we may as well apply here. The CC problem is a mater of debate, no doubt about it, but once more we should concur with him that it is above all a matter of observation; for observing if $\CC$ is a rigid parameter or a dynamical variable can also greatly help in the way we should finally tackle the CC problem! It may actually lead to the clue for its understanding. Is this not, after all, how real physics works? It is time for less formal theory and more observations!}. It is our contention that a better understanding of $H_0$ is related to a deeper knowledge of the nature of $\CC$, in particular whether the rigid option $\CC=$const. could be superseded by the more flexible notion of dynamical vacuum energy density. Obviously this has an implication on the value of $H_0$, as shown in Table 1.

The actual value of $H_0$ has a long and tortuous history, and the tension among the different measurements is inherent to it. Let us only recall that after Baade's revision (by a factor of one half\,\cite{Baade1944}) of the exceedingly large value $\sim 500$ km/s/Mpc originally estimated by Hubble (implying a universe of barely two billion years old only), the Hubble parameter was further lowered to $75$ km/s/Mpc and finally was captured in the range $H_0=55\pm 5$ km/s/Mpc, where it remained for 20 years (until 1995), mainly under the influence of Sandage's devoted observations\,\cite{Tammann1996}. See nevertheless \cite{vandenBergh1996} for an alternative historical point of view, in which a higher range of values is advocated.
Subsequently, with the first historical observations of the accelerated expansion of the universe, suggesting a positive value of $\CC$\,\cite{SNIaRiess,SNIaPerl}, the typical range for $H_0$ moved upwards to $\sim 65$ km/s/Mpc. In the meantime, a wealth of observational values of $H_0$ have piled up in the literature using different methods (see e.g. the median statistical analysis of $>550$ measurements considered in \cite{ChenRatra2011,BethapudiDesai2017}).

\begin{figure}
\begin{center}
\label{contours}
\includegraphics[width=2.7in]{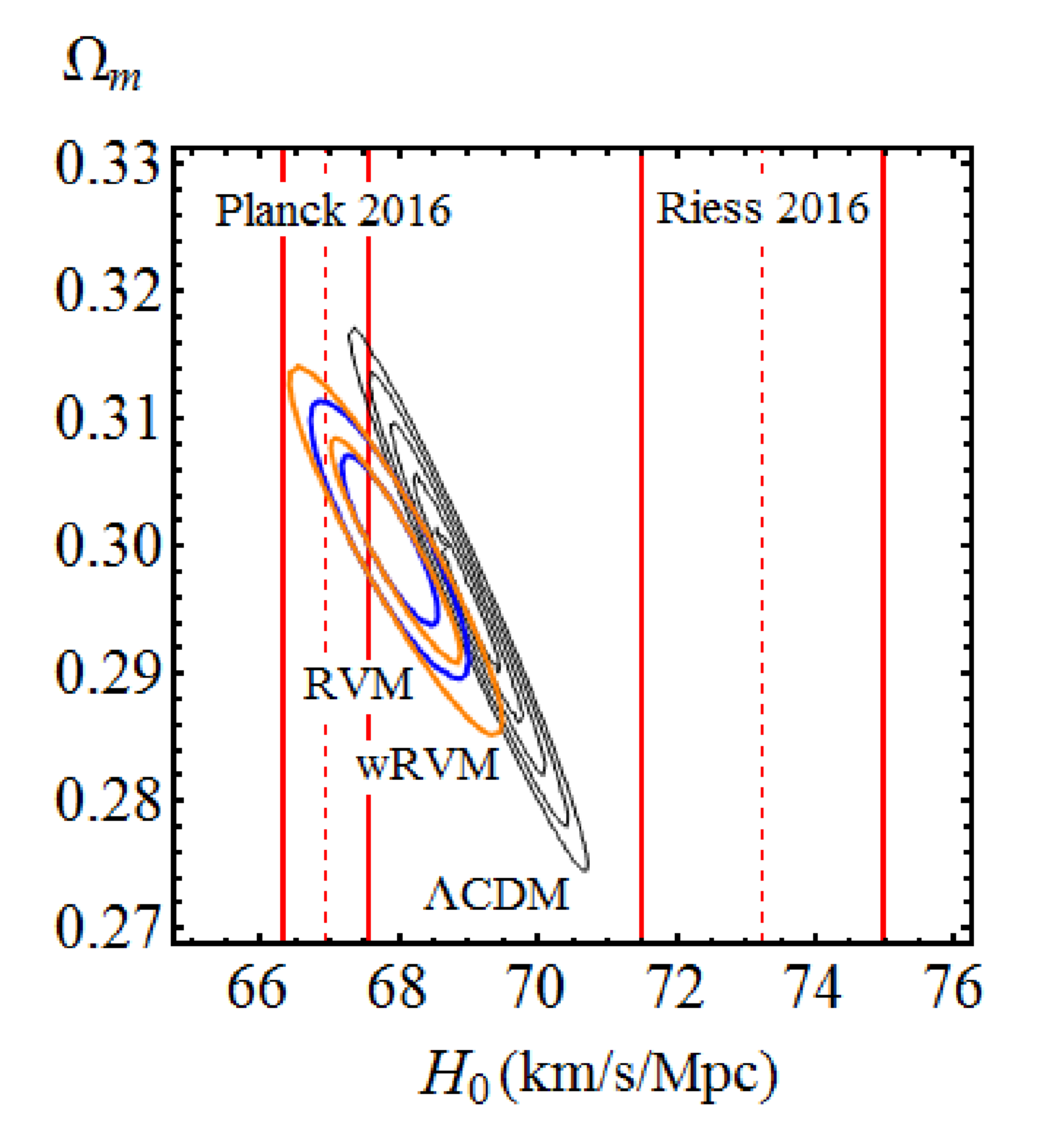}
\caption{\scriptsize Contour plots in the $(H_0,\Omo)$-plane for the RVM (blue) and $w$RVM (orange) up to $2\sigma$, together with those for the $\CC$CDM (black) up to $5\sigma$, corresponding to the situation when the local $H_0$ value of Riess et al.\,\cite{RiessH02016} is included as a data point in the fit (cf. Table 2) \,\cite{PLB2017}.
}
\end{center}
\end{figure}

\subsection{The documented tension}

As mentioned above, two kinds of \emph{precision} (few percent level) measurements of $H_0$ have generated considerable perplexity in the recent literature, specifically between the latest Planck values ($H^{\rm Planck}_0$) obtained from the CMB anisotropies, and the local HST measurement (based on distance estimates from Cepheids). The latter, obtained by Riess et al.\,\cite{RiessH02016}, is
\begin{equation}\label{eq:H0Riess}
H_0^{\rm Riess}= 73.24\pm 1.74\ {\rm km/s/Mpc}
\end{equation}
whereas the CMB value, depending on the kind of data used, reads\cite{Planck2015}
\begin{equation}\label{eq:H0CMB2015}
H^{\rm Planck\, 2015}_0= 67.51\pm 0.64\ {\rm km/s/Mpc}\ ({\rm TT,TE,EE+lowP}+{\rm lensing\ data})\,,
\end{equation}
and\cite{Planck2016}
\begin{equation}\label{eq:H0CMB2016}
H^{\rm Planck\, 2016}_0= 66.93\pm 0.62\ {\rm km/s/Mpc}\ ({\rm TT,TE,EE+SIMlow\ data} )\,.
\end{equation}
In both cases there is a tension above $3\sigma$ c.l. (viz. $3.1\sigma$ and $3.4\sigma$, respectively) with respect to the local measurement\,\footnote{It is suggestive to see that the late local measurements of $H_0$ obtained by Sandage and Tammann\,\cite{TammannSandage2010}, based on Cepheids and SNIa, are closer to the current CMB measurements than to the recent Riess et al. local measurements based on similar techniques. It makes one think if the current local measurement is actually an outlier.}. We will refer the Planck measurement collectively as $H_0^{\rm Planck}$ since the tension with the local measurement (\ref{eq:H0Riess}) is essentially the same.  This situation, and in general a certain level of tension with some independent observations
in intermediate cosmological scales, has stimulated a number of discussions and possible solutions in the literature, see e.g.\,\cite{Melchiorri2016,Bernal2016,Shafieloo2017,Cardona2017,Melchiorri2017a,Melchiorri2017b}.

\subsection{Refiting the overall data in the presence of $H_0^{\rm Riess}$}

We wish to reexamine here the $H^{\rm Riess}_0-H^{\rm Planck}_0$ tension, but not as an isolated conflict between two particular sources of observations, but rather in light of the overall fit to the current cosmological  SNIa+BAO+$H(z)$+LSS+CMB data. In other words, it is worthwhile to reconsider the fitting results of Table 1 when we introduce $H_0^{\rm Riess}$ as an explicit data point in the fit. The result is recorded in Table 2.

\begin{figure}
\begin{center}
\label{contours}
\includegraphics[width=5in]{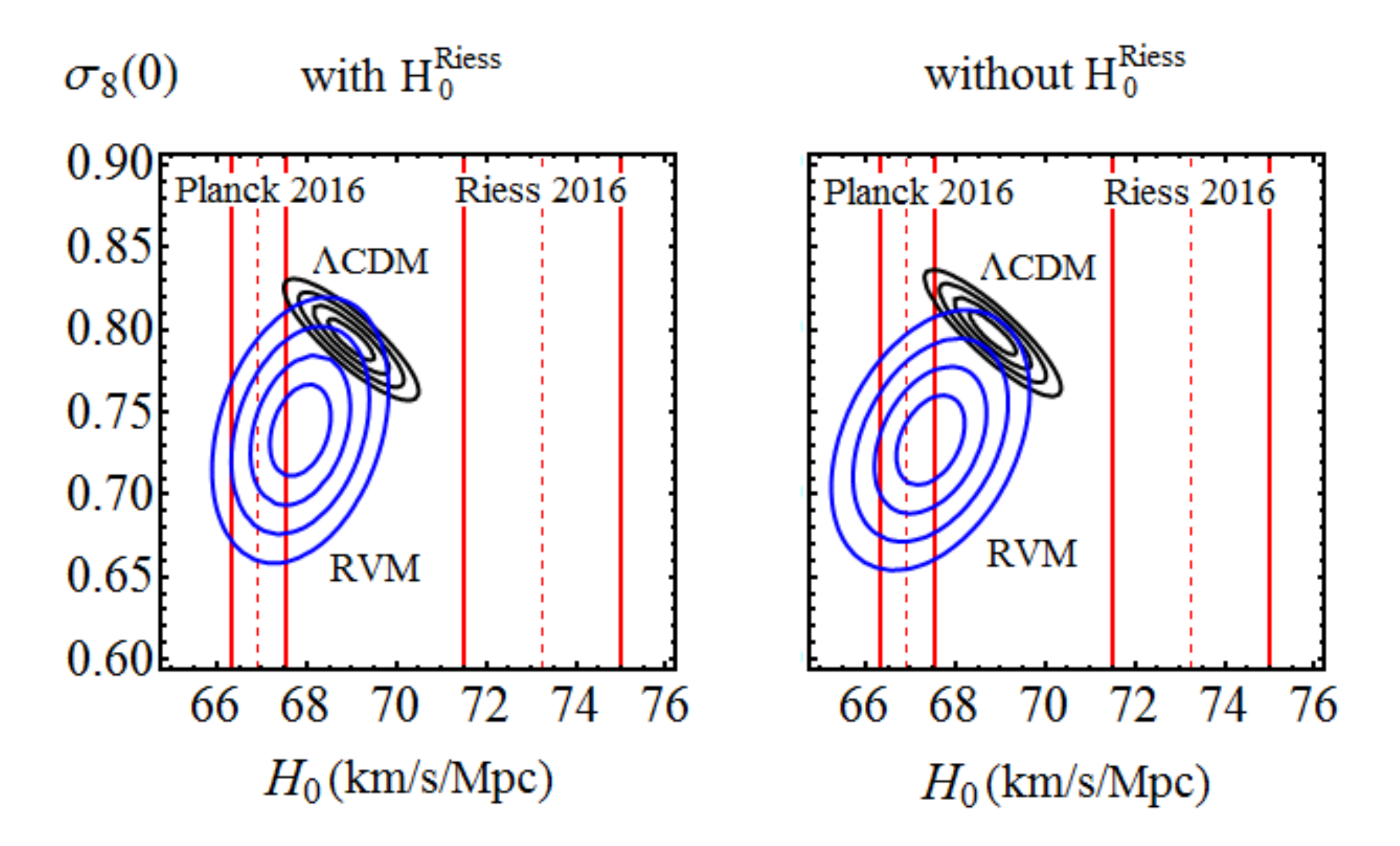}
\caption{\scriptsize \newtext{Contour lines for the $\CC$CDM (black) and RVM (blue) up to $4\sigma$ in the $(H_0,\sigma_8(0))$-plane. We present in the {\it left plot} the case when the local $H_0$ value of Riess {\it et al.}\,\cite{RiessH02016} is included in the fit (cf. Table 2), whereas in the {\it right plot} the case when that local value is {\it not} included (cf. Table 1). See \cite{PLB2017} for more details.}}
\end{center}
\end{figure}

In Fig. 2 we plot  $f(z)\sigma_8(z)$ for the various models I, II and III for the case with constant $w\neq-1$ using the fitted values of Table 2  and some other special situations described in the caption. The case $w=-1$ is not plotted in Fig.\,2 because it is visually undistinguishable from the case $w$ near $-1$. The numerical differences in the fitting results, however, are not negligible as can be see on comparing Tables 1 and 2.

In Fig.\,3 we display the contour plots in the $(H_0,\Omo)$-plane for the RVM (blue) and $w$RVM (orange) up to $2\sigma$, together with those for the $\CC$CDM (black) up to $5\sigma$, corresponding to the situation when the local $H_0$ value of Riess et al.\,\cite{RiessH02016} is included as a data point in the fit (cf. Table 2) \,\cite{PLB2017}. One can see that when all data sources SNIa+BAO+$H(z)$+LSS+CMB are used, the price for reaching the vicinity of $H_0^{\rm Riess}$ is a too small value of $\Omo$ around $0.27$ and requires extended contours beyond  $5\sigma$ c.l. The figure also shows that both the RVM and $w$RVM intersect much better (already at $1\sigma$) the $H_0^{\rm Planck}$ range than the $\CC$CDM. The latter requires also $5\sigma$ contours to reach $H^{\rm Planck}_0$, and $\Omo$ near $0.32$. In other words, when the local value $H^{\rm Riess}_0$ enters the fit,  the  $\CC$CDM is in a rather uncomfortable position, as it is almost far-equidistant from both the $H_0^{\rm Planck}$ and $H^{\rm Riess}_0$
domains!

Finally, in Fig.\, 4  we show once more the difficulty  of reaching the $H_0^{\rm Riess}$ neighborhood, as it  enforces to extend the contours beyond the $5\sigma$ c.l., what would imply  a too low value of $\Omega_m$ in both cases (cf. Fig. 3) and, in addition, would result in a too large value of $\sigma_8(0)$ for the RVM.  Interestingly,  $H_0$ and $\sigma_8(0)$ are positively correlated in the RVM (i.e. $H_0$ decreases when $\sigma_8(0)$ decreases), whilst they are anticorrelated in the $\CC$CDM ($H_0$ increases when $\sigma_8(0)$ decreases, and vice versa). It is this opposite correlation feature with respect to the $\CC$CDM what allows the RVM to improve the LSS fit in the region where both $H_0$ and $\sigma_8(0)$ are smaller than the respective $\CC$CDM values (cf. Fig. 4). This explains why the Planck range for $H_0$ is clearly preferred by the RVM, as it allows a much better description of the LSS data.

While we must still remain open to the possibility that the $H^{\rm Planck}_0$ and/or $H^{\rm Riess}_0$ measurements are affected by some kind of (unknown) systematic errors, some of these possibilities may be on the way of  being ruled out by recent works. In \cite{Aylor2017} the authors study the systematic errors in Planck's data by comparing them with the South Pole Telescope data. They conclude  that there is no evidence of systematic errors in Planck's results.  Let us also mention the ``blinded''  determination  $H_0 = 72.5\pm 3.2$ km/s/Mpc  from \cite{Zhang2017}, based on a reanalysis of the SNIa and Cepheid variables data from the older work by Riess et al., where it was found  $H_0 = 73.8\pm 2.4$ km/s/Mpc\,\cite{RiessH02011}.  The tension with $H^{\rm Planck}_0$ diminishes, compare with Eq.\,(\ref{eq:H0Riess}),  since the central value decreased and  the uncertainty has grown significantly by more than $\sim 30\%$.

On the other hand, in  \cite{Addison2017}  it is shown that by combining the latest BAO results with WMAP, Atacama Cosmology Telescope (ACT), or South Pole Telescope (SPT) CMB data produces values of $H_0$  that are $2.4-3.1\sigma$ lower than the distance ladder, independent of Planck. These authors conclude from their analysis  that it is not possible to explain the $H_0$  disagreement solely with a systematic error specific to the Planck data. In another vein the work\,\cite{Follin2017} excludes systematic bias or uncertainty in the Cepheid calibration step of the distance ladder measurement by\,\cite{RiessH02016}. Finally, we mention the  recent study \cite{Lin2017}, in which the authors run a new (dis)cordance test based on  using a recently introduced index of inconsistency  capable of dissecting inconsistencies between two or more data sets.  After comparing the constraints on $H_0$ from different methods and observing the decreasing behavior of such index when the local $H_0$-measurement is removed they conclude that such local
measurement is an outlier compared to the others, what would favor a systematics-based explanation. This  is compatible with the observed improvement in the statistical quality of our analysis when the local $H_0$-measurement is removed from our overall fit.

The search for a final solution to the $H_0$ tension is, of course, still work in progress.  The class of the $(w)$RVMs studied here offers a viable solution to both the $H_0$ and $\sigma_8(0)$ existing tensions in the data, which are both unaccountable within the $\CC$CDM.  Thus, within our  dynamical vacuum framework the CMB-based Planck value of $H_0$  is definitely favored and is compatible with a smaller value of $\sigma_8(0)$,  whereas the local measurement of $H_0$  is considered an outlier, namely a value which seems to be irreconcilable with a simultaneous solution of the other $\CC$CDM tensions.

\section{Conclusions}
 In this work, we have reviewed the status of the dynamical vacuum models (DVMs) in their ability to compete with the $\CC$CDM model (namely the standard or concordance model of cosmology) to fit the overall SNIa+BAO+$H(z)$+LSS+CMB cosmological observations. We find that the the current cosmological data disfavors the $\CC$CDM, and hence the  $\CC=$const. hypothesis, in a very significant way.
The best fit value to the overall data is provided by the running vacuum model (RVM), at a confidence level of roughly $\sim 4\sigma$ as compared to the $\CC$CDM. Even a simple XCDM parametrization yields roughly $3\sigma$.  These results are consistent with our most recent studies\,\cite{ApJL2015,ApJ2017,MPLA2017,PRD2017}.

We have also used these models to reanalyze the tension between the Riess et al. local measurement $H_0^{\rm Riess}$ and the value obtained in the CMB measurements from the Planck satellite, $H_0^{\rm Planck}$, which is $3\sigma$ smaller.  We find that the fit quality to the overall SNIa+BAO+$H(z)$+LSS+CMB  cosmological data increases to the maximum level only when the local $H_0^{\rm Riess}$ measurement is not taken into account. In other words, the CMB determination of $H_0$ is clearly preferred. We demonstrate that not only the CMB and BAO, but also the LSS data, are essential to grant these results.

We have also comparatively considered the performance of the $w$DVMs (i.e. the dynamical quasi-vacuum models with $w\neq -1$), and we have found that they are also able to improve the $\CC$CDM fit, although to a lesser extent than the best DVMs.  However, the extra degree of freedom associated to the free parameter $w$ in these models can be used to try to enforce a minimal $H^{\rm Riess}_0-H^{\rm Planck}_0$ tension. What we find is that if the LSS data are not considered in the fit analysis, the tension can indeed be diminished, but only at the expense of a phantom-like dynamical behavior of the DE, namely $w$ turns out to satisfy $w\lesssim -1$.

But the main problem is that this implies a serious disagreement with the structure formation data. Such disagreement disappears when the LSS data are restored, and in fact a good fit quality to the overall observations (better than the $\CC$CDM at $3\sigma$ c.l.) can be achieved, with no trace of phantom dynamical DE energy.  In the absence of the  local  $H_0^{\rm Riess}$ measurement, the fit quality in favor of the main dynamical vacuum models further increases up to $\sim4\sigma$ c.l. In general the vacuum dynamics tends to favor the CMB determination of $H_0$ against the local measurement $H_0^{\rm Riess}$, but this measurement can still be accommodated in the fit without seriously spoiling the capacity of the RVM to improve the $\CC$CDM fit.

These results are bolstered by outstanding marks of the information criteria (yielding values $\Delta$AIC$>10$ {\it and} $\Delta$BIC$>10$) in favor of the main DVMs and against the concordance model. To summarize, we claim that significant signs of dynamical vacuum energy density are sitting in the current cosmological data, which the concordance $\CC$CDM model is unable to accommodate.

\section{Acknowledgments}
J. Sol\`a is thankful to Prof. Harald Fritzsch for the kind invitation to this stimulating conference on Cosmology, Gravitational Waves and Particles.  He would also like to thank Prof. K. K. Phua for inviting him to present this contribution in the review
section of IJMPA. We have been supported by MINECO FPA2016-76005-C2-1-P, Consolider CSD2007-00042, 2014-SGR-104 (Generalitat de Catalunya) and MDM- 2014-0369 (ICCUB). J. Sol\`a is also particularly grateful for the support received from the Institute for Advanced Study of the Nanyang Technological University in Singapore, where part of this work was performed.

\vspace{0.7cm}
\noindent{\bf \Large Note Added:}
\newline
{Since the first version of this work appeared, new analyses of the cosmological data have been published, in particular the one-year results by the DES collaboration (DES Y1 for short)\,\cite{DES2017}. The  Bayes factor indicates that the DES Y1 and Planck data sets are consistent with each other in the context of $\CC$CDM and therefore both are insensitive to any effect on dynamical DE.  In the case of  DES Y1  they do not use direct $f(z)\sigma_8(z)$  data on LSS structure formation despite they recognize that smaller values of $\sigma_8(0)$ than those predicted by the $\CC$CDM are necessary to solve the tension between the concordance model and the LSS observations.

Another recent work that does not find any sign of dynamical DE is \cite{Heavens2017}. As we have explained in \,\cite{PRD2017}, we attribute this lack of sensitivity  once more to the fact of  not using large scale structure formation data and BAO. The latter is indeed not used  in that paper e.g. for the analysis of DE  in terms of the XCDM parametrization, and they do not use data on $f(z)\sigma_8(z)$ anywhere in their analysis.  We contend that by restricting to  CMB and lensing data is not sufficient to be sensitive to a possible evolution of the DE.

This is also the reason why Planck did not report any such evidence.
In previous works of us-- see  in particular \,\cite{ApJ2017,PRD2017} -- we have explored quantitatively in detail why is so. For instance, in \cite{Planck2015} the Planck team did not use  LSS (RSD) data at all, and in \cite{PlanckDE2015} they  used a rather restricted set of  BAO and LSS points. In our previous works  \,\cite{ApJ2017,PRD2017} we have explicitly demonstrated that under the same limited conditions used by Planck  (not different from those of DES Y1 in this respect)  we do recover their negative results on dynamical DE.

However, when we use the full data string, which involves not only CMB but also the rich  BAO+LSS data set that has been described in \,\cite{ApJ2017,PRD2017}, then we obtain substantially positive indications of dynamical DE at a confidence level in between $3-4\sigma$.  While incontestable evidence requires of course  $5\sigma$ c.l. at least,  these results are already quite encouraging. In fact, they are actually consistent with the recent analysis by Gong-Bo Zhao et al.\,\cite{GongBoZhao2017}, who reported on a signal of dynamical DE at $3.5\sigma$ c.l using similar data ingredients as in our analysis. See also the complementary comments on these results provided in \cite{DiValentino2017}, where it is suggested  that  they seem to point to a ``crack in the current cosmological paradigm''.}


\newcommand{\CQG}[3]{{ Class. Quant. Grav. } {\bf #1} (#2) {#3}}
\newcommand{\JCAP}[3]{{ JCAP} {\bf#1} (#2)  {#3}}
\newcommand{\APJ}[3]{{ Astrophys. J. } {\bf #1} (#2)  {#3}}
\newcommand{\AMJ}[3]{{ Astronom. J. } {\bf #1} (#2)  {#3}}
\newcommand{\APP}[3]{{ Astropart. Phys. } {\bf #1} (#2)  {#3}}
\newcommand{\AAP}[3]{{ Astron. Astrophys. } {\bf #1} (#2)  {#3}}
\newcommand{\MNRAS}[3]{{ Mon. Not. Roy. Astron. Soc.} {\bf #1} (#2)  {#3}}
\newcommand{\PR}[3]{{ Phys. Rep. } {\bf #1} (#2)  {#3}}
\newcommand{\RMP}[3]{{ Rev. Mod. Phys. } {\bf #1} (#2)  {#3}}
\newcommand{\JPA}[3]{{ J. Phys. A: Math. Theor.} {\bf #1} (#2)  {#3}}
\newcommand{\ProgS}[3]{{ Prog. Theor. Phys. Supp.} {\bf #1} (#2)  {#3}}
\newcommand{\APJS}[3]{{ Astrophys. J. Supl.} {\bf #1} (#2)  {#3}}

\newcommand{\Prog}[3]{{ Prog. Theor. Phys.} {\bf #1}  (#2) {#3}}
\newcommand{\IJMPA}[3]{{ Int. J. of Mod. Phys. A} {\bf #1}  {(#2)} {#3}}
\newcommand{\IJMPD}[3]{{ Int. J. of Mod. Phys. D} {\bf #1}  {(#2)} {#3}}
\newcommand{\GRG}[3]{{ Gen. Rel. Grav.} {\bf #1}  {(#2)} {#3}}



\end{document}